\documentclass[preprin,showpacs,preprintnumbers,eqsecnum,amsmath,amssymb,nofootinbib]{revtex4}
\usepackage{graphics,epsfig,xcolor}
\usepackage{graphicx}
\usepackage [latin1]{inputenc}

\begin{document}
\baselineskip=0.8 cm
\title{$D$-dimensional aether charged black hole and aether waves  in  M-subclass of Einstein-aether theory}
\author{Chikun Ding$^{1}$}\thanks{Email: dingchikun@163.com}
\author{Yuebing Zhou$^{1}$}\author{Yu Shi$^{1}$}\author{Xiangyun Fu$^{2}$}
\affiliation{$^1$Department of Physics, Huaihua University, Huaihua, 418008, P. R. China
\\ $^2$Institute of Physics, Hunan University of Science and Technology, Xiangtan, Hunan 411201, P. R. China}
\vspace*{0.2cm}
\begin{abstract}
\baselineskip=0.6 cm

\begin{center}{\bf Abstract}\end{center}
We obtain an exact $D$-dimensional aether charged black hole solution and gravitational wave polarizations in the M-subclass of the Einstein-aether theory with Lorentz invariance violated by an unit norm vector field---the aether field $u^\mu $. This aether field can be timelike or spacelike and, the aether charge $Q_{\ae}$ has a nonzero minimum value, which is different from the electric charge. The aether electric-like potential $u_t$ is regular, and the aether magnetic-like potential $u_r$ is singular at the horizons. Though the Lorentz symmetry is broken, the Smarr formula and the first law of black hole thermodynamics can be exactly constructed via the extended method of Killing potential. However, we find this conception of the aether charge doesn't exist in the $c_i$ subclass of the Einstein-aether theory. For the linearized M-subclass Einstein-aether theory with the timelike aether field, we find the speed of spin-2 modes (the usual gravitational wave) is still equal to 1. But there is only one polarization $\gamma_{12}$ that can propagate, while the other one, $\gamma_{11}$ cannot propagate for $D=4$ due to the effect of Lorentz symmetry breaking. The speed of spin-1 modes (the transverse aether wave) is also equal to 1. The third kind mode is the longitudinal aether-metric mode, which is linearly time dependent and not the spin-0 mode reported in the $c_i$ subclass Einstein-aether theory.\end{abstract}

 \maketitle
\vspace*{0.2cm}
\section{Introduction}
Gravity, described by General Relativity (GR), has been the most difficult force to bring under theoretical control at the classical and quantum levels. In GR, the presence of a dimensional gravity coupling constant, $G=m^{-2}_{pl}$, leads to the non-renormalizability as a quantum field theory in four dimensional spacetime. One possible way to evade this problem is to introduce a constant norm vector $u^\mu $, known as aether field \cite{gasperini,jacobson}, to break the fundamental symmetry---Lorentz invariance. This is because that the Lorentz group is noncompact and the exact Lorentz invariance will lead to divergences in quantum field theory associated with states of arbitrarily high energy and momentum \cite{jacobson}. Einstein-aether gravitational theory is a modification of GR
where the kinematic quantities of an unitary timelike aether field $u^\mu $ are introduced in the gravitational action integral by the form of a Lagrange multiplier potential $\lambda(u^\mu u_\mu-1)$. This aether field has an unit norm and timelike everywhere indicating the preferred  frame which will break the boost sector of the Lorentz symmetry in the whole spacetime. Such a field carries a nonlinear representation of the local Lorentz group and can be called a theory of nonlinearly  realized Lorentz invariance. The preferred vector \textcolor[rgb]{0.00,0.00,1.00}{is} not necessarily timelike \cite{kostelecky198940}, so the aether field can also be an unitary spacelike vector which will violate the rotation sector of the Lorentz symmetry with the Lagrange multiplier potential $\lambda(u^\mu u_\mu+1)$ used in early universe \cite{dulaney}, for deriving a black hole solution \cite{dimakis}, for studying the extra dimensions \cite{carroll,chatrabhuti}, for studying thin accretion disks \cite{he}, etc. This aether field also can be set to null, $u^\mu u_\mu=0$, which enables one to naturally introduce a scalar degree of freedom \cite{gurses2019,gurses201979}. These aether models are nothing but vector having a non-vanishing vacuum expectation in the Higgs mechanism \cite{griffiths} and serve as a phenomenological representation of the Lorentz violating terms in the gravitational sector of the Standard Model Extension \cite{pelaggi}.

To maintain general covariance of the theory, the preferred threading is required to be dynamical, so Einstein-aether theory is distinguished from other Lorentz-violating theories restoring Lorentz invariance at low energy scales. The most general form of the covariant Lagrangian of Einstein-aether theory is \cite{jacobson}
\begin{eqnarray}\label{aether}
&&\mathcal{L}=
a_0-a_1R-c_1(\nabla_\mu u_\nu)(\nabla^\mu u^\nu)-c_2(\nabla_\mu u^\mu )^2
-c_3(\nabla_\mu u_\nu)(\nabla^\nu u^\mu)-c_4a^2
+\lambda(u^2\pm1),
\end{eqnarray}
where $R$ is Ricci scalar, $a^k=u^m\nabla_mu^k$ is the aether acceleration, $\lambda$ is a Lagrange multiplier that enforce the unit constraint,  $``\pm"$ signs represent the spacelike and timelike aether field \footnote{Under the metric signs of $(+,-,\cdots,-)$, $u^\mu $ is timelike when $u^\mu u_\mu=1$; spacelike when $u^\mu u_\mu=-1$.}, respectively.
The most extensively researched subclass of it   contains four nonzero coupling constants $c_1,c_2,c_3,c_4$ (named as ``$c_i$-subclass" Einstein-aether theory) besides the Ricci scalar $R$ and the Lagrange multiplier potential $\lambda$ term with a timelike aether field. The exact static black hole solutions within this subclass were found by Berglund {\it et al} \cite{berglund}, the numerical static solutions that satisfied the self-consistent conditions were found by Zhang and Wang {\it et al} \cite{zhang}, the exact charged static solutions were found by Ding and Lin {\it et al} \cite{ding2015,ding2016,ding2019,lin}, the even- and odd-parity perturbations on a static aether black hole were studied by Mukohyama and Felice {\it et al} \cite{mukohyama,felice}. In this $c_i$-subclass aether theory, the introduce of an unit timelike aether vector allows some particles can travel faster than the speed of light, and an universal horizon exists behind the Killing horizon that can trap excitations traveling at arbitrarily high velocities.

Due to the $c_2$ term $(\nabla_\mu u^\mu )^2$ in the action (\ref{action}) is equivalent to the combination of terms $(\nabla_\mu u_\nu)(\nabla^\mu u^\nu)-(1/2) F^{\mu\nu}F_{\mu\nu}+R_{\mu\nu}u^\mu u^\nu$ \cite{jacobson}, where $F_{\mu\nu}=\nabla_\mu u_\nu -\nabla_\nu u_\mu$ is the aether field strength, it is interesting that the $c_i$-subclass Einstein-aether theory can be reduced to the subclass of $\mathcal{L}_2=-a_1R-b_1F^{\mu\nu}F_{\mu\nu}-c_2(\nabla_\mu u^\mu )^2+\lambda(u^2\pm1)$  by the redefinition of the metric and the aether field \cite{foster}, where $b_1$ is a positive constant. In 2013, Gao {\it et al} derived a black hole solution in this subclass with a timelike aether field  \cite{gao}. This solution is similar to the Reissner-Nordstrom solution but not asymptotically flat. Recently, G\"urses and Hajian {\it et al} studied the subclass of  $\mathcal{L}_3=-a_1R-c_2(\nabla_\mu u^\mu )^2+\lambda(u^2\pm1)$ \cite{gurses,hajian}. They showed that an effective cosmological constant can be reduced as a result of local symmetry breaking induced by a non-zero vacuum value of the aether field. Jacobson and Kostelecky {\it et al} \cite{jacobson,kostelecky198940} studied the subclass of the nonzero constants $\mathcal{L}_4=-a_1R-b_1F^{\mu\nu}F_{\mu\nu}+\lambda(u^2\pm1)$ which comprises a subclass of the solutions to the coupled Einstein-Maxwell-charged dust equations, which can be named as ``M-subclass" Einstein-aether theory. In Ref. \cite{kostelecky198940}, the authors showed that this model  could not provide masses for components of the graviton via gravitational version of the Higgs effect. In Ref. \cite{jacobson}, the authors showed that this model included all vacuum solutions of general relativity with timelike aether field $u_\mu=(u_t,u_r,0,0)$. However, they didn't give the high dimensional case, nor the Lorentz violation effect on the gravitational wave.

The motivations for study on higher than four dimensional black holes are as following. If one uses GR  to describe the geometry of spacetime, then it should not be restricted only to 4-dimensions \cite{pereniguez}. Endowing GR with a tunable parameter $D$ should lead to valuable insight into the nature of the theory \cite{emparan}. The string theory contains gravity and requires more than four dimensions. Brane World model showed us that our four dimension universe is actually a membrane embedded in a five dimensional spacetime. The AdS/CFT correspondence relates the dynamics of a $D$-dimensional black hole with those of a quantum field theory in $D-1$ dimensions without gravity. In the large $D$ limit, recent study shows that black holes behave as non-interacting particles \cite{emparan2013}.

So we will study the $D$-dimensional aether charged black hole solution and aether waves in this M-subclass of Einstein-aether gravity theory, try to find the effect of the Lorentz invariance violation on the conventional black hole and gravitational wave physics, and the difference from the $c_i$ Einstein-aether theory.  The rest of the paper is organized as follows.   In Sec. II we give the background for the M-subclass of Einstein-aether theory and derive the gravitational field equations, aether motion equations. In Sec. III, we derive  black hole solutions, give the form of aether charge $Q_{\ae}$ and corresponding potential $V_{\ae}$, and construct the Smarr formula and the first law of black hole thermodynamics for the aether charged black hole and find the effect of Lorentz violation on black hole thermodynamics. Sec. IV studies the linearized equations for the gravitational and aether waves. Sec. V gives a summary. The appendix shows some components of Einstein-tensor and momentum tensor.

\section{M-subclass of Einstein-aether gravity theory}

In the Einstein-aether gravity model (\ref{aether}), one introduces the dynamical vector field $u_{a}$ which has a nonzero vacuum expectation value, leading to a spontaneous Lorentz symmetry breaking in the gravitational sector via a given potential.
 The action of the M-subclass Einstein-aether theory in the $D$ dimensional spacetime is \cite{kostelecky198940},
\begin{eqnarray}
\mathcal{S}=
\int d^Dx\sqrt{-g}\Big[-\frac{R}{2\kappa} -\frac{b_1}{4}F^{\mu\nu}F_{\mu\nu}
+\lambda(u_\mu u^{\mu}\mp1)\Big], \label{action}
\end{eqnarray}
where $b_1$ is a positive constant\footnote{This assumption is to guarantee that the aether charge contribution has the standard Reissner-Nordstrom like form.}, and $``-"$ sign represents the timelike aether and $``+"$ sign for the spacelike aether.  $\kappa=8\pi G_D/c^4$, where $G_D$ is the $D$ dimensional gravitational constant. Here and hereafter, we take $G_D=1$ and $c=1$, and the metric signs of $(+,-,\cdots,-)$ for convenience.

Varying the action (\ref{action}) with respect to the metric yields the gravitational field equations
\begin{eqnarray}\label{einstein0}
G_{\mu\nu}=\kappa T_{\mu\nu}^{\ae},
\end{eqnarray}
where the Einstein's tensor $G_{\mu\nu}=R_{\mu\nu}-g_{\mu\nu}R/2$,
the aether energy momentum tensor $T_{\mu\nu}^{\ae}$ is
\begin{eqnarray}\label{momentum}
&&T_{\mu\nu}^{\ae}=-b_1(F_{\mu\alpha}F^{\alpha}_{\;\nu}-\frac{1}{4}g_{\mu\nu} F^{\alpha\beta}F_{\alpha\beta})+g_{\mu\nu}\lambda(u^2\mp1)+
2\lambda u_{\mu}u_{\nu}.
\end{eqnarray}
Varying instead with respect to the aether field and Lagrange multiplier $\lambda$, it generates the aether equations of motion and constraint,
\begin{eqnarray}\label{motion}
\nabla ^{\mu}F_{\mu\nu}=-\frac{2\lambda}{b_1} u_\nu,\qquad u^2=\pm1.
\end{eqnarray}
This constraint means $u^\mu $ is an unit timelike vector and/or spacelike vector\footnote{Eqs. (\ref{momentum}) and (\ref{motion}) are similar to those of Einstein-Maxwell-charged dust system: $2\lambda u_\mu u_\nu$ corresponding to $\rho u_au_b$, $-2\lambda u_\nu/b_1$ corresponding to $4\pi \rho_eu_a$ \cite{lemos}, where $u_a$ is the four-velocity of the dust fluid, $\rho$ is the density of the charged dust, $\rho_e$ is its electric charge density. But here the aether field cannot be treated as charged dust, because there is actually no electromagnetic field at all. Here we can treated it as an aether dust.}.
The contracted Bianchi identities ($\nabla ^\mu G_{\mu\nu}=0$) lead to conservation of the total energy-momentum tensor
\begin{eqnarray}\label{}
\nabla ^\mu T^{\ae}_{\mu\nu}=0.
\end{eqnarray}

The static spherically symmetric black hole metric has the form\begin{eqnarray}\label{metric}
&&ds^2=e^{2\phi(r)}dt^2-e^{2\psi(r)}dr^2-r^2d\Omega^2_{D-2},
\end{eqnarray}
where $\Omega_{D-2}$ is a standard $(D-2)$-sphere.
In this static spherically symmetric spacetime, the most general form for the aether field would be $u_\mu=(u_t,u_r,0,0,\cdots,0)$, where $u_t$ and $u_r$ are functions of $r$ subject to the constraint $u_t^2 e^{ -2\phi}- u_r^2 e^{-2 \psi} =\pm1$\footnote{This constraint requires that the square root to be real globally for all $r>0$.}. In this general case, the aether field has both radial and a time component for the vacuum expectation value. So the aether field is supposed to be timelike or spacelike  as that
\begin{eqnarray}\label{bu}
u_\mu=\big(h(r),e^{\psi(r)}\sqrt{e^{-2\phi(r)}h^2(r)\mp1},0,0,\cdots,0\big).
\end{eqnarray}

To solve the gravitational field equation (\ref{einstein0}), we list the nonzero components of Einstein tensor and the aether energy momentum tensor in the Appendix.
One can find that there aren't the cross-term components of Einstein tensor $ G_{01},$ $ G_{10}$. However, there are non-zero components of the aether energy momentum tensor  $T_{01}^{\ae}$, $T_{10}^{\ae}$. So from the field equation (\ref{einstein0}), the terms $T_{01}^{\ae}$ and $T_{10}^{\ae}$ should be vanishing, which leads to $\lambda=0$. In Ref. \cite{jacobson}, the authors also pointed out that there aren't static solutions for $\lambda\neq0$.  From Eq. (\ref{momentum}), $\lambda=0$ means that here the aether field looks more like an electromagnetic field with no source. But here $u_{\mu}$ is actually not the electromagnetic potential, it has an unit norm, timelike or spacelike everywhere. Therefore, we will study this source-free Maxwell-like branch with $\lambda=0$.

One can obtain the following three  component equations for Eq. (\ref{einstein0}),
\begin{eqnarray}
&&\frac{(D-2)e^{2\phi}}{2r^2}\big[(D-3)(e^{2\psi}-1)+2r\psi'\big]=\frac{\kappa}{2}b_1h'^2,\label{tt}\\
&&\frac{(D-2)e^{2\phi}}{2r^2}\big[(D-3)(1-e^{2\psi})+2r\phi'\big]=-\frac{\kappa}{2}b_1h'^2,\label{rr}\\
&&e^{2\phi}\Big[\frac{(D-3)(D-4)}{2r^2}(1-e^{2\psi})+\frac{(D-3)}{r}(\phi'-\psi')+\phi''-\phi'\psi'+\phi'^2\Big]=\frac{\kappa}{2}b_1h'^2
,\label{theta}
\end{eqnarray}
and Eq. (\ref{motion}) gives the aether field motion equation,
\begin{eqnarray}h'(-D+2+r\phi'+r\psi')-rh''=0,\label{hpr}
\end{eqnarray}
 the prime $'$ is the derivative with respect to the corresponding argument, respectively.
Adding the Eq. (\ref{tt}) to (\ref{rr}), one can obtain that
\begin{eqnarray}\label{}
\phi'+\psi'=0.\label{theta2}
\end{eqnarray}
Substituting it to Eq. (\ref{hpr}), one can obtain
 \begin{eqnarray}
h(r)=\frac{4\pi Q_{\ae}}{(D-3)V_{D-2}r^{D-3}}+C_1,\label{hr}
\end{eqnarray} where $Q_{\ae}$ is a positive integral constant indicating the aether charge, $C_1$ is the second integral constant which can be determined via the boundary conditions. $V_{D-2}$ is the volume of the $(D-2)$-sphere, $V_{D-2}=2\sqrt{\pi}^{(D-1)}/\Gamma[(D-1)/2]$ \cite{ding2023}. Here for convenient from Eq. (\ref{theta2}), we let $e^{2\phi}=A(r)$ and $e^{2\psi}=1/A(r)$, so the Eq. (\ref{tt}) can give that,
\begin{eqnarray}\label{metricar}
(D-2)\big[r A'(r)+(D-3)A(r)-(D-3)\big]+\frac{\kappa b_1 (4\pi Q_{\ae})^2}{V^2_{D-2}r^{2(D-3)}}=0,
\end{eqnarray}
and Eq. (\ref{bu}) becomes
\begin{eqnarray}\label{}
u_\mu=\big(h(r),\frac{\sqrt{h^2(r)\mp A(r)}}{A(r)},0,0,\cdots,0\big).
\end{eqnarray}
Here we impose the asymptotically flat conditions that the metric function $A(r)\rightarrow1$ at infinity $r\rightarrow\infty$. While for the timelike aether field, the components $u_t\rightarrow1$ and $u_r\rightarrow0$; for the spacelike aether field, $u_t\rightarrow0$ and $u_r\rightarrow1$ at infinity. Under the above boundary conditions, the integral constant $C_1$ can be fixed, i.e., when the aether field is timelike, the function $h(r)$ is,
 \begin{eqnarray}
h(r)=\frac{4\pi Q_{\ae}}{(D-3)V_{D-2}r^{D-3}}+1;\label{hrt}
\end{eqnarray}
when the aether field is spacelike, the function $h(r)$ is,
 \begin{eqnarray}
h(r)=\frac{4\pi Q_{\ae}}{(D-3)V_{D-2}r^{D-3}}.\label{hrs}
\end{eqnarray}
Then the aether field strength $F_{\mu\nu}=\nabla_\mu u_\nu-\nabla_\nu u_\mu$ is,
 \begin{eqnarray}
F_{01}=-F_{10}=\frac{4\pi Q_{\ae}}{V_{D-2}r^{D-2}},\label{}
\end{eqnarray}
which is the same as that of the electromagnetical field strength.
In the next section, we will derive the solution $A(r)$ and study its thermodynamical properties.

\section{Aether charged black hole solutions and Smarr formula}

In this section, we study the aether charged black hole solution, and try to find the effect of Lorentz violation on a black hole physics. From the Eq. (\ref{metricar}), one can obtain that,
\begin{eqnarray}\label{metricc}
A(r)=1-\frac{2\tilde{M}}{r^{D-3}}+\frac{\tilde{Q}_{\ae}^2}{r^{2(D-3)}}
\end{eqnarray}
with
\begin{eqnarray}\label{mandq}
\tilde{M}=\frac{8\pi M}{(D-2)V_{D-2}},\qquad \tilde{Q}_{\ae}^2=\frac{\kappa b_1(4\pi Q_{\ae})^2}{(D-2)(D-3)V^2_{D-2}},
\end{eqnarray}where $M$ is the mass of the black hole.
The metric (\ref{metric}) becomes,
\begin{eqnarray}\label{metric1}
&&ds^2=A(r)dt^2-\frac{1}{A(r)}dr^2-r^2d\Omega^2_{D-2}.
\end{eqnarray}
This black hole solution has the form of $D$-dimensional Reissner-Nordstrom black hole when the aether coupling constant $b_1=2/\kappa=1/4\pi$ \cite{myers,wei,wang,cai}, and $M, Q_{\ae}$ in Eqs. (\ref{mandq}) are the conserved charges in this aether charged black hole spacetime. The difference is that $Q_{\ae}$ is the aether charge, not the electric charge; and the aether field has a constant norm everywhere, but the norm of the electric field $\bar{A}_\mu \bar{A}^\mu=q^2/r^2A(r)$ isn't a constant, where $\bar{A}_\mu$ is the electromagnetic potential, $\bar{A}_\mu=(q/r,0,0,0)$ with the electric charge $q$. In Ref. \cite{gao}, the aether potentials $u_t $ and $u_r$  correspond to the electric and magnetic part of the electromagnetic potential, i.e., $u_t$ is treated as the electric potential, $u_r$ as the magnetic potential. So the aether field here can be treated as an electromagnetic-like field.

This aether charged black hole  looks also like the black hole on the brane \cite{dadhich}, where there is no electric charge but instead a tidal charge, arising from the projection onto the brane of free gravitational field effects in the bulk.
When the aether field is timelike, $u^\mu u_\mu=1$, it has the form of,
\begin{eqnarray}\label{}
u_\mu=\Big(h, \frac{\sqrt{h^2-A(r)}}{A(r)}, 0, \cdots,0\Big),\;\; u^\mu=\Big(\frac{h}{ A(r)}, -\sqrt{h^2-A(r)}, 0, \cdots, 0\Big),
\end{eqnarray}
where $h(r)$ has the form of Eq. (\ref{hrt}), and $h^2-A(r)\geq0$ means that $0<\kappa b_1\leq(D-2)/(D-3)$, which shows that the aether coupling constant $b_1$ is limited by the spacetime dimensions $D$. One can see that at the infinity $r\rightarrow\infty$, the aether electric-like potential(temporal component of the aether field) $u_t\rightarrow1$, the aether magnetic-like potential (radial component of the aether field) $u_r\rightarrow0$ \footnote{Note that in $c_i$ subclass Einstein-aether theory, the timelike aether field $u_\mu\rightarrow(1,0,\cdots,0)$ \cite{ding2019} at infinity.}. Their behaviors with different $Q_{\ae}$ are showed in Fig. 1 when $D=4$.
 \begin{figure}[ht]
\begin{center}
\includegraphics[width=5.0cm]{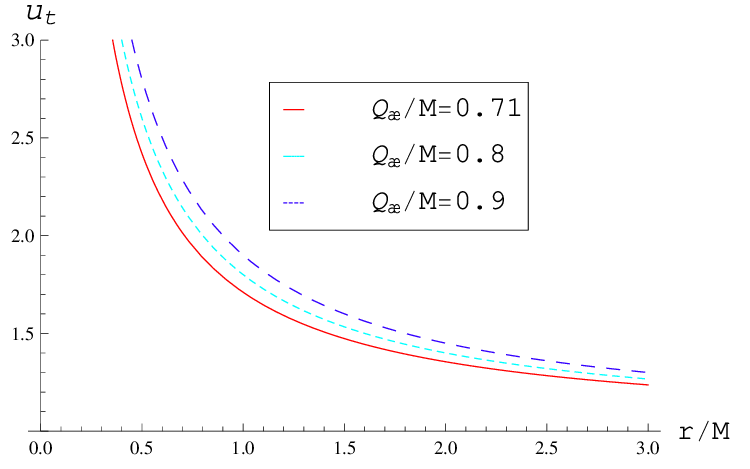}\;\;\;\;\includegraphics[width=5.0cm]{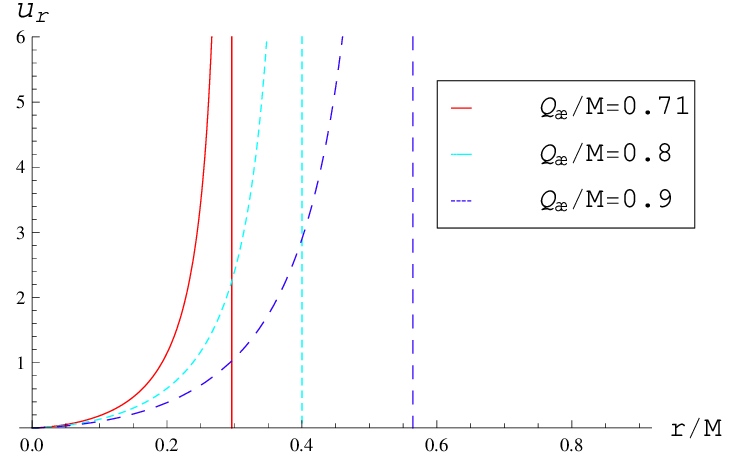}
\;\;\;\;\includegraphics[width=5.0cm]{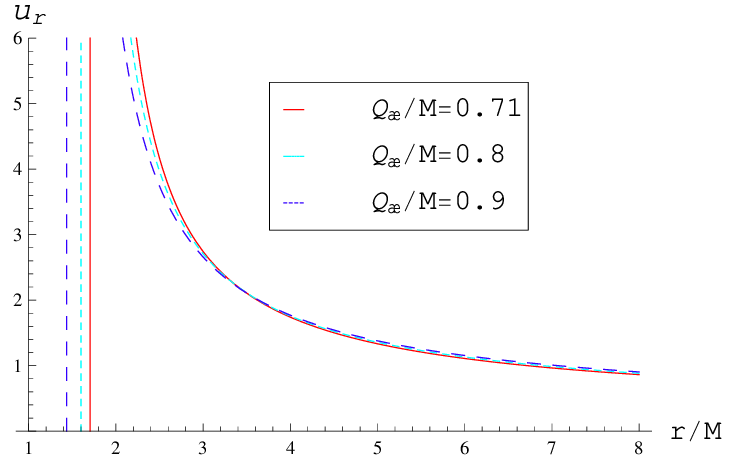}
\caption{The variation of the aether electric-like potential $u_t$ and magnetic-like potential $u_r$ of the aether charged black hole with different aether charge $Q_{\ae}$ when $D=4,b_1=2/\kappa=1/4\pi$ for timelike aether field. Here the aether charge $Q_{\ae}$ satisfies the condition that $Q_{\ae}\leq M$ and the coupling constant $b_1$ satisfies the condition that  $0<b_1\leq 2/\kappa$. One can see that $u_t$ is regular, while $u_r$ is singular at the horizons $r_h/M=1\pm\sqrt{1-Q_{\ae}^2/M^2}$.}\label{fp1}
 \end{center}
 \end{figure}

When the aether field is spacelike, $u^\mu u_\mu=-1$, it has the form of,
\begin{eqnarray}\label{}
u_\mu=\Big(h, \frac{\sqrt{h^2+A(r)}}{A(r)}, 0, \cdots,0\Big),\;\; u^\mu=\Big(\frac{h}{ A(r)}, -\sqrt{h^2+A(r)}, 0, \cdots, 0\Big),
\end{eqnarray}
where $h(r)$ has the form of Eq. (\ref{hrs}), and $h^2+A(r)\geq0$ means that,
\begin{eqnarray}\label{}
\tilde Q_{\ae}\geq\sqrt{\frac{\kappa b_1(D-3)}{(D-2)+\kappa b_1(D-3)}}\tilde M.
\end{eqnarray}
It shows that when the aether field is spacelike, the aether charge has a nonzero minimum value, which is different from the electric charge. One can see that at the infinite, $u_r\rightarrow1$, or $u_\mu\rightarrow(0,1,0,\cdots,0)$. At the origin $r\rightarrow0$, the aether electric-like potential is singular, $u_t\rightarrow\infty$; but the aether magnetic-like potential is regular, $u_r\rightarrow0$. At the horizon $r_h$,  the aether electric-like potential is regular, but the aether magnetic-like potential is singular, $u_r\rightarrow\infty$. $u_t$ and $u_r$ satisfy the condition that $u_t^2/A(r)-A(r)u^2_r=-1$. These behaviors are showed in Fig. 2 when $D=4$.
 \begin{figure}[ht]
\begin{center}
\includegraphics[width=5.0cm]{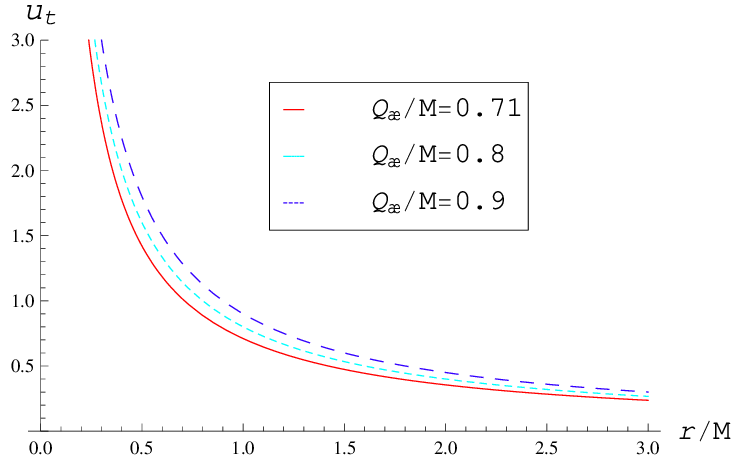}\;\;\;\;\includegraphics[width=5.0cm]{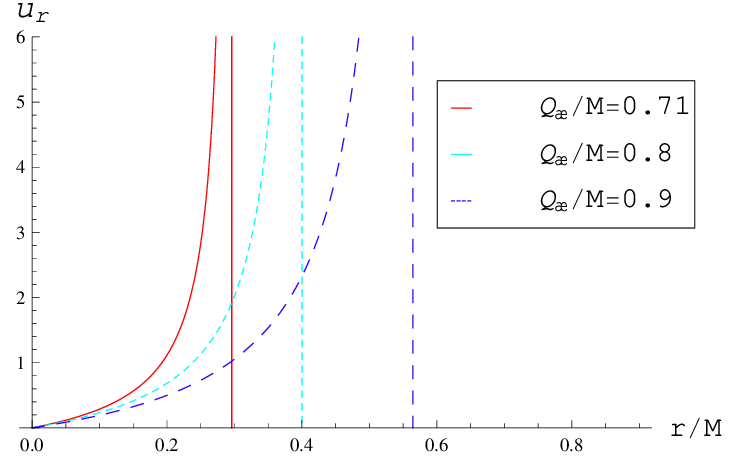}
\;\;\;\;\includegraphics[width=5.0cm]{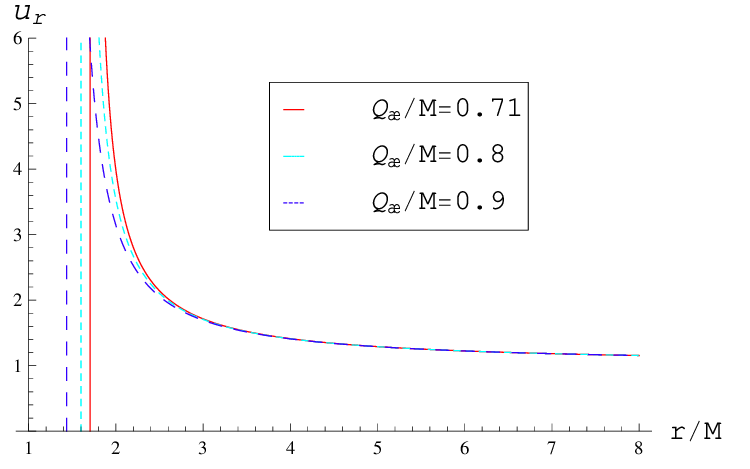}
\caption{The variation of the aether electric-like potential $u_t$ and magnetic-like potential $u_r$ of the aether charged black hole with different aether charge $Q_{\ae}$ when $D=4,b_1=2/\kappa=1/4\pi$ for the spacelike aether field. Here the aether charge $Q_{\ae}$ satisfies the condition that $M/\sqrt{2}\leq Q_{\ae}\leq M$. One can see that $u_t$ is regular, while $u_r$ is singular at the horizons $r_h/M=1\pm\sqrt{1-Q_{\ae}^2/M^2}$.}\label{fp2}
 \end{center}
 \end{figure}

The Killing horizons of this aether charged black hole are,
\begin{eqnarray}\label{horizon}
r_h^{D-3}=\tilde{M}\pm\sqrt{\tilde{M}^2-\tilde Q_{\ae}^2},
\end{eqnarray}
requiring that the aether charge should be less than the black hole mass, $\tilde Q_{\ae}\leq \tilde{M}$. However, there is no universal horizon reported in Refs. \cite{berglund,ding2015} for the $c_i$ subclass Einstein-aether theory. Its Hawking temperature is
\begin{eqnarray}\label{}
T=\frac{D-3}{4\pi r_h}\left[1-\frac{\tilde Q_{\ae}^2}{r_h^{2(D-3)}}\right].
\end{eqnarray}

Now, we derive the Smarr formula by introducing Killing potential $\omega^{\mu\nu}$ and construct the first law of black hole thermodynamics for the aether charged  black hole. Suppose that $\mathcal{M}$ is this $D$-dimensional spacetime satisfying the Einstein equations, $\xi^\mu=(1,0,\cdots,0)$ is a Killing vector on $\mathcal{M}$, timelike near infinity. In $\mathcal{M}$,  there is a spacelike hypersurface $\tilde{S}$ with a dimension $(D-2)$-surface boundary $\partial \tilde{S}$, and $\xi^\mu$ is normal to the $\tilde{S}$. The boundary $\partial \tilde{S}$ has two components: an inner boundary at the event horizon $\partial \tilde{S}_h$ and an outer boundary at infinity $\partial \tilde{S}_\infty$.
We can integrate the Killing equation $\nabla_\nu(\nabla^\nu\xi^\mu)=-R^\mu_\nu\xi^\nu$ over this hypersurface $\tilde{S}$,
\begin{eqnarray}\label{intone}
\int_{\partial \tilde{S}}\nabla^\nu\xi^\mu d\Sigma_{\mu\nu}=-\int_{\tilde{S}}R^\mu_\nu\xi^\nu d\Sigma_\mu,
\end{eqnarray}
where $d\Sigma_{\mu\nu}$ and $d\Sigma_{\mu}$ are the surface elements of $\partial \tilde{S}$ and $\tilde{S}$, respectively.
The non-vanishing components of the tensor $\nabla^\nu\xi^\mu$ and $R_t^t$ are given by
\begin{eqnarray}\label{Rtt}
\nabla^r\xi^t=-\nabla^t\xi^r=-\frac{A'(r)}{2},\quad R_t^t=\frac{(D-3)^2\tilde Q_{\ae}^2}{r^{2(D-2)}}.
\end{eqnarray}
Since Ricci tensor $R^\mu _\nu$ is nonzero, Gauss's law cannot be used to the right side of Eq. (\ref{intone}). However in Ref.  \cite{kastor}, Komar integral relation still holds in the $D$-dimensional  dS/AdS black hole spacetime with $R_{t}^t=\Lambda \neq0$, by introducing an anti-symmetric Killing potential $\omega^{\mu\nu}$ which can be obtained according to relation $\xi^\nu=\nabla_\mu\omega^{\mu\nu}$.
In Ref. \cite{ding2023}, we used it to study thermodynamics for the high dimensional dS/AdS bumblebee black hole. However for the aether charged  black hole, there is no this cosmological constant ``$\Lambda$", but $R^\mu_\nu$ is a $r$ dependent function. Therefore the above method need to be extended \cite{ding2025}. For the present static Killing vector $\xi^\mu$, we have nonzero components of $\omega^{\mu\nu}$ that,
\begin{eqnarray}\label{}
\omega^{rt}=-\omega^{tr}=\frac{r}{D-1}+ \frac{\alpha}{r^{2(D-3)}},
\end{eqnarray}
where $\alpha$ is an integral constant. Due to $R_t^t$ in Eq. (\ref{Rtt}) is not a constant but a function of $r$, one should
introduce another function $g(r)$ to lead to the relation that,
\begin{eqnarray}\label{}
\nabla_\nu[g(r)\omega^{\mu\nu}]=R^t_t\nabla_\nu\omega^{\mu\nu}.
\end{eqnarray}
We find that,
\begin{eqnarray}\label{}
g(r)=\frac{(D-1)(D-3)\tilde{Q}^2_{\ae}}{r^{2(D-2)}}.
\end{eqnarray}
Lastly, the Eq. (\ref{intone}) can be rewritten as
\begin{eqnarray}\label{inttwo}
\frac{D-2}{\kappa}\int_{\partial \tilde{S}}\left(\nabla^\nu\xi^\mu+g(r)\omega^{\mu\nu}\right)d\Sigma_{\mu\nu}=0,
\end{eqnarray}
which is multiplied by the normalization factor $(D-2)/\kappa$ and known as the Komar integral relation.

The closed 2-surface $\partial \tilde{S}$ has two parts, horizon $\partial \tilde{S}_h$ and infinite $\partial \tilde{S}_\infty$,
so Eq. (\ref{inttwo}) can be rewritten as
\begin{eqnarray}\label{intthree}
\frac{D-2}{\kappa}\int_{\partial \tilde{S}_\infty}\left[\nabla^\nu\xi^\mu+g(r)\omega^{\mu\nu}\right]d\Sigma_{\mu\nu}=\frac{D-2}{\kappa}\int_{\partial \tilde{S}_h}\left[\nabla^\nu\xi^\mu+g(r)\omega^{\mu\nu}\right]d\Sigma_{\mu\nu}.
\end{eqnarray}
With the $(D-2)$-surface element $d\Sigma_{rt}=-r^{D-2}d\Omega_{D-2}/2$, the left and right hand sides of this integral are,
 \begin{eqnarray}\label{}
&&\frac{D-2}{\kappa}\int_{\partial \tilde{S}_\infty}\left[\nabla^\nu\xi^\mu+g(r)\omega^{\mu\nu}\right]d\Sigma_{\mu\nu}
=(D-3)M,\\&& \frac{D-2}{\kappa}\int_{\partial \tilde{S}_h}\left[\nabla^\nu\xi^\mu+g(r)\omega^{\mu\nu}\right]d\Sigma_{\mu\nu}
=(D-2)TS+(D-3)V_{\ae}Q_{\ae},
\end{eqnarray}
where the aether charge potential $V_{\ae}$ is
\begin{eqnarray}\label{smar}
V_{\ae}=\frac{16\pi^2b_1Q_{\ae}}{(D-3)V_{D-2}r_h^{D-3}},
\end{eqnarray}
which is approximate to (not the same as) the value of the aether electrical potential $u_t$ at the horizon.
Lastly, the integral (\ref{intthree}) can give the following Smarr formula,
\begin{eqnarray}\label{smar}
(D-3)M=(D-2)TS+(D-3)V_{\ae} Q_{\ae},
\end{eqnarray}
which is the same as that in Refs. \cite{cvetic,kubiznak} when the aether coupling constant $b_1=2/\kappa=1/4\pi$, where the entropy $S$ still follows the area law, i.e., one quarter of the horizon area $\tilde{A}$ \cite{rodrigues},
\begin{eqnarray}\label{}
4S=\tilde{A}=\int_{horizon}r_h^{D-2}d\Omega_{D-2}=r_h^{D-2}V_{D-2}.
\end{eqnarray}
This is because that the aether field does not destruct the covariance of the action (\ref{action}).

The first law of black hole thermodynamics can be constructed by the following method as,
\begin{eqnarray}\label{}
dM=\left(\frac{\partial M}{\partial S}\right)_{Q_{\ae}}dS+\left(\frac{\partial M}{\partial Q_{\ae}}\right)_SdQ_{\ae}.
\end{eqnarray}
From the black hole's horizon Eq. (\ref{horizon}), one can rewrite the black hole mass as following,
\begin{eqnarray}\label{}
M=\frac{(D-2)V_{D-2}}{16\pi}\left[\left(\frac{4S}{V_{D-2}}\right)^{\frac{D-3}{D-2}}
+\tilde Q_{\ae}^2\left(\frac{V_{D-2}}{4S}\right)^{\frac{D-3}{D-2}}\right].
\end{eqnarray}
Lastly, the first law is,
\begin{eqnarray}\label{}
dM=TdS+V_{\ae}dQ_{\ae}.
\end{eqnarray}
One can see that the conventional Smarr formula and the first law of black hole physics still hold for this Lorentz violation aether charged black hole. However, in $c_i$ subclass Einstein-aether theory, the first law cannot be constructed exactly \cite{ding2019}, where the form of the aether field also is $u_\mu=\big(u_t(r),u_r(r),0,\cdots,0\big)$, but there is no conception of aether charge \footnote{In  $c_i$ subclass Einstein-aether theory when $c_{14}=0$, the black hole solution is \cite{berglund,ding2017} \begin{eqnarray}A(r)=1-\frac{2M}{r}-\frac{c_{13}r^4_{\ae}}{r^4}.\nonumber\end{eqnarray} When $c_{123}=0$,  the black hole solution is \begin{eqnarray}A(r)=1-\frac{2M}{r}-\frac{r_u^2}{r^2}.\nonumber\end{eqnarray}  However, the quantity $r_{\ae}$ or $r_{u}$ is not an independent parameter, it is $r^4_{\ae}=27M^4/16(1-c_{13})$, or $r_u^2=(c_{13}-c_{14}/2)M^2/(1-c_{13})$. Therefore, there is no conception of aether charge.}.

\section{Polarizations of the gravitational and aether waves}

In this section, we study the linearized M-subclass Einstein-aether theory in higher $D$-dimensional spacetime and find the speeds and polarizations of all the wave modes with the timelike aether field. The first step is to linearize the field equations about the flat background solution with Minkowski metric and constant unit vector $\b u^\mu $. The fields are expanded as
\begin{eqnarray}
g_{\mu\nu}=\eta_{\mu\nu}+\gamma_{\mu\nu},\;u^\mu =\b u^\mu +v^\mu,\;\lambda=0+\lambda^{(1)}
\end{eqnarray}
where $\eta_{\mu\nu}=$diag$(1,-1,\cdots,-1)$  for $D$-dimensional flat Minkowski spacetime $(x^0,x^i)$ and $\b u^\mu =(1,0,\cdots,0)$, $ (i=1,2,\cdots,d)$, $d=D-1$. $\gamma_{\mu\nu}, v_\mu, \lambda^{(1)}$ are small quantities. Note that the background solution of the Lagrange multiplier $\lambda$ is zero for the flat spacetime.  In this section we use the flat metric $\eta_{\mu\nu}$ to raise and lower indices.

Keeping only the first order terms in $v^\mu$ and $\gamma_{\mu\nu}$, the field equations (\ref{einstein0}, \ref{motion}) become (timelike aether field $u^\mu u_\mu=1$),
\begin{eqnarray}\label{line}
G_{\mu\nu}^{(1)}=2\kappa \lambda^{(1)}\b u_\mu\b u_\nu,\;\;\partial_\mu F^{(1)\mu}_{~~~~~m}=-\frac{2\lambda^{(1)}}{b_1} \b u_m,\;\;v^0+\frac{1}{2}\gamma_{00}=0,
\end{eqnarray}
where the superscript ``(1)" denotes the first order part of the corresponding quantity, and
\begin{eqnarray}\label{gab}
G_{\mu\nu}^{(1)}=\frac{1}{2}\Box\gamma_{\mu\nu}
+\frac{1}{2}\gamma_{,\mu\nu}-\gamma_{m(\mu,\nu)}^{\quad\quad m}-\frac{1}{2}\eta _{\mu\nu}(\Box\gamma-\gamma_{mn,}^{\quad mn}),\;\partial_\mu F^{(1)\mu}_{~~~~~m}=\Box v_m-\partial_m(\partial^\mu v_\mu),
\end{eqnarray}
where $\gamma=\gamma^{\;m}_m$.
We can use Lorentz gauge conditions $\bar \gamma_{\mu\nu}^{\;\;\;\;,\nu}=0,\; \partial_\mu v^\mu=0$ to simplify the Eqs. (\ref{line}) and(\ref{gab}) as following, where $\bar \gamma_{\mu\nu}=\gamma_{\mu\nu}-\eta_{\mu\nu}\gamma/2$,
\begin{eqnarray}
&&
\Box\big(\gamma_{\mu\nu}-\frac{1}{2}\eta_{\mu\nu}\gamma\big)=4\kappa \lambda^{(1)}\b u_\mu\b u_\nu,\label{boxg}\\&&
\Box V_{\ae}=-\frac{2\lambda^{(1)}}{b_1} \b u_\nu,\label{boxv}
\end{eqnarray}
which have some wavelike solutions. In Eqs. (\ref{boxg}) and (\ref{boxv}), $2\lambda^{(1)}$ can be treated as the density of the aether dust, $-2\lambda^{(1)}/b_1$ corresponds to the density of the aether charge.

 At first, we consider the aether wave Eq. (\ref{boxv}). Obviously, there are equations that $\Box v_i=0$ and the aether perturbations has a form of plane wave solution that \begin{eqnarray}\label{}
v_i=\epsilon_ie^{ik_cx^c},
\end{eqnarray}
and the coordinates are chosen such that the wave vector is $k_c=(k_0,0,\cdots,0,k_d)$, where $d=D-1$, $\epsilon_i$ is the vectorial wave module, $i=1,2,\cdots, d-1$.  The residual gauge freedom allows one to impose $v_d=0$ due to that $\Box v_d=0$. However, $\Box v_0\neq0$ and so $v_0$ cannot be imposed to zero. Here we have to set $\partial_0v_0=0$, only in this way, the condition $\partial^\mu v_\mu=0$ can still hold, i.e., $v_0$ is a time independent nonzero function.

Nextly, we consider the gravitational wave Eq. (\ref{boxg}). Obviously, there are equations that $\Box \gamma_{kl}=0$ and the metric perturbations has a form of plane wave solution that \begin{eqnarray}\label{}
\gamma_{kl}=\epsilon_{kl}e^{ik_cx^c},
\end{eqnarray}
where $k\neq l$ and $k,l=1,2,...,d-1$, and $\epsilon_{kl}$ is the tensorial wave module. The residual gauge freedom allows one to impose $\gamma_{0i}=0$ and $\gamma_{kd}=0$ due to that they satisfy the wave equations. However, $\gamma$ and $\gamma_{dd}$ don't satisfy the wave equation due to the source term. Therefore, we cannot set $\gamma=0$ and $\gamma_{dd}=0$ as usual in GR. Because $v_0$ is time independent, then $\gamma_{00}=-2v^0$ is also time independent. The diagonal components of $\gamma_{ij}$ satisfy the following equation,
\begin{eqnarray}\label{}
\Box(\gamma_{11}+\gamma_{22}+...+\gamma_{dd})+\frac{d}{2}\Box\gamma=0,
\end{eqnarray}
which allows one to set $\gamma=-2\gamma_{00}/(d-2)$. So from Eq. (\ref{boxg}), the diagonal components are
\begin{eqnarray}\label{traceless}
\gamma_{11}=\gamma_{22}=...=\gamma_{dd}=\frac{1}{d-2}\gamma_{00},
\end{eqnarray}
which are all time independent. From the above equation, one can see that the trace $\gamma_{11}+\gamma_{22}+...+\gamma_{(d-1)(d-1)}\ne0$, i.e., the traceless metric modes $\gamma_{11}+\gamma_{22}+...+\gamma_{(d-1)(d-1)}=0$ in GR \cite{ding2019} disappear! At the same time we add the limitation that $\gamma_{dd,d}=0$, then the above gauge condition $\bar \gamma_{\mu\nu}^{\;\; ,\nu}=0$ can still hold.

Since one cannot set $v_0$ to zero, then there must exist the longitudinal part of the aether wave $v_k^L=\partial_kf$ which is related to $v_0$, where $f$ is a scalar field. The 00-component of the Eqs. (\ref{boxg}) and (\ref{boxv}) are,
\begin{eqnarray}
\partial^k\partial_k\gamma_{00}=\frac{d-2}{d-1}4\kappa \lambda^{(1)},\qquad \partial^k\partial_kv_{0}-\partial_0(\partial^k\partial_kf)=-\frac{2\lambda^{(1)}}{b_1}.
\end{eqnarray}
Solving the above both equations, one can obtain that,
\begin{eqnarray}
f=\hat{A}v_0(x^k)t+B(x^k),\qquad \hat{A}=\left[1-\frac{d-1}{b_1\kappa(d-2)}\right],
\end{eqnarray}
or,
\begin{eqnarray}
f=-\frac{\hat{A}}{2}\gamma_{00}(x^k)t+B(x^k),
\end{eqnarray}
where $B(x^k)$ is an arbitrary function of spatial variable $x^k,\;k=1,2,...,(d-1)$. Then the longitudinal aether wave is $v_k^L=\partial_kf$, which is linear time dependent. This linear time dependence mode corresponds to the linearization of the restricted gauge symmetry under the gauge transformation $u'_\mu\rightarrow u_\mu+\nabla_\mu\varphi$. It belongs to the redundant degrees of freedom in linearized gravity \cite{jacobson}, where  all non-linear terms of the second order and higher are completely neglected and artificially breaks the self-consistent coupling between the gravitational field and the dust matter---aehter field. This anomaly mode is completely non-physical redundancy and does not correspond to any real physical propagation mode.

Lastly, we can conclude that there are three kinds of wave modes. The first is the transverse $\epsilon_{kl}\neq0$ modes, corresponding to the usual gravitational waves in GR when polarization components $\epsilon_{k}$ vanish (unexcited aether waves). The speed of these usual gravitational wave modes---spin-2,
\begin{eqnarray}\label{}
s^2_2=1,
\end{eqnarray}
which is the same as that in 4-dimensional spacetime indicating that the gravitational wave modes spin-2 aren't dependent on spacetime dimensions and aether constant $b_1$. This result also is the same as that in the $c_i$-subclass Einstein-aether theory when the constants $c_{13}=0$ \cite{ding2019}. However, their polarizations are suffered by the aether field---when $D=4$, there is only one polarization $\gamma_{12}$ propagating, but the other---traceless metric mode $\gamma_{11}$ or $\gamma_{22}\;(\gamma_{11}+\gamma_{22}=0)$ disappears (from Eq. (\ref{traceless}), $\gamma_{11}=\gamma_{22}=\gamma_{00}/2=-v_0$, which is time independent and cannot propagate)!

The second kind modes correspond  to transverse aether when polarizations $\epsilon_{k}$ is nonzero. In four dimensional spacetime, there are two modes: $v_1$ and $v_2$.  The speed of the transverse aether wave modes---spin-1
\begin{eqnarray}\label{}
s^2_1=1,
\end{eqnarray}
which is the same as that in the $c_i$-subclass Einstein-aether theory when the constants $c_{13}=0$ and $c_4=0$ \cite{ding2019}. It also aren't dependent on spacetime dimensions and aether constant $b_1$. The spin-1 modes speed is equivalent to spin-2 modes.

The third kind mode are trace aether-metric corresponding to nonzero polarization $\gamma_{00}$ which is dependent on the dimension number $D$. In four dimensional spacetime, it is
\begin{eqnarray}\label{}
v_k^L=-\frac{\hat{A}t}{2}\partial_k\gamma_{00}(x^1,x^2)+\partial_kB(x^1,x^2),
\end{eqnarray}
where $k=1,2$. It is because that the aether dust energy density $2\kappa\lambda^{(1)}$ is adjusted to produce an arbitrary gravitational potential $\gamma_{00}$ which is time independent. This purely static gravitational potential perturbation can directly drive aether dust particles to acquire a velocity that increases linearly with time via the linearized geodesic equation. Therefore, the third kind mode isn't the spin-0 mode appeared in Ref. \cite{ding2019} for the $c_i$ subclass Einstein-aether theory.

\section{Summary}

In this paper we study the $D$-dimensional black hole solution and gravitational wave polarizations in a  M-subclass of the Einstein-aether theory with Lorentz invariance violated by an unit norm vector field---the aether field $u_\mu=\big(u_t(r),u_r(r),0,\cdots,0\big)$. This M-subclass of Einstein-aether theory has a form of Einstein-Maxwell theory with a term of Lagrange multiplier  potential $\lambda(u^\mu u_\mu\mp1)$. We find that when $\lambda=0$, there exists a static solution---$D$-dimensional Reissner-Nordstrom-like black hole solution which is aether charged, not electrically charged. If  the aether field is timelike, the aether charge is less than black hole mass $\tilde Q_{\ae}\leq\tilde{M}$ and the coupling constant $0<\kappa b_1\leq(D-2)/(D-3)$. If the aether field is spacelike,  the aether charge has a nonzero minimum value, which is different from the electric charge.  This aether charge is bounded by the black hole mass, \begin{eqnarray}\label{}
\sqrt{\frac{\kappa b_1(D-3)}{(D-2)+\kappa b_1(D-3)}}\tilde M\leq \tilde Q_{\ae}\leq\tilde{M},
\end{eqnarray}
but the coupling constant $b_1$ isn't constrained.
The form of the components $u_t, u_r$ are also different: if $u_\mu$ is timelike, it is the Eq. (\ref{hrt}); if spacelike, it is the  Eq. (\ref{hrs}). This conception of aether charge doesn't exist in the $c_i$ subclass of Einstein-aether theory.

The aether field $u_\mu$ is timelike or spacelike in the whole spacetime, and its components, the aether electric potential $u_t$ is regular, while the aether magnetic potential $u_r$ is singular at the horizons, and they satisfy the condition that $u_t^2/A(r)-A(r)u^2_r=\pm1$. Though the Lorentz invariance has been violated by the aether field, the  Smarr formula and the first law of black hole physics can still be exactly constructed by using the extended method of the Killing potential. This means that the local Lorentz violation does not affect the black hole thermodynamics in the M-subclass of Einstein-aether theory.

For the linearized M-subclass Einstein-aether theory with the timelike aether field, we find the speed of spin-2 modes (the usual gravitational wave) is still equal to 1, which aren't dependent on spacetime dimensions $D$ and aether constant $b_1$, but the traceless metric  polarizations are disappeared. For $D=4$, there is only one gravitational wave polarization $\gamma_{12}$, the other, $\gamma_{11}$, or $\gamma_{22}=-\gamma_{11}$ cannot propagate and disappear due to the whole spacetime timelike aether field $u^\mu u_\mu=1$. This is the effect of the Lorentz symmetry breaking on the usual gravitational waves in GR.

The speed of spin-1 modes (the aether transverse wave) is also equal to 1. The speed of the above both kinds modes are the same as those in $c_i$ subclass Einstein-aether theory when $c_{13}=c_4=0$ \cite{ding2019}. The third kind mode is the longitudinal aether-metric mode, which is dependent on the spacetime dimensions. However, this mode is linearly time dependent, and its time derivative is arbitrary. Therefore it is no the spin-0 mode reported in Ref. \cite{ding2019} for the $c_i$ subclass Einstein-aether theory. These anomaly longitudinal modes  also exist in other theories, such as $f(R)$ theory \cite{gogoi}, Horndeski theory \cite{hou}.

\begin{acknowledgments}  Zhou's work was supported by the National Natural Science Foundation of China(NNSFC) under  Grant No. 12375047; Fu's work was supported by NNSFC under  Grant No. 12375045;
Shi's work was supported by NNSFC under  Grant No. 12305063; Guo's work was supported by Key Project Research of the Education Department of Hunnan Province, No. 25A0575.
\end{acknowledgments}

\appendix
\section{Some nonzero components of the given tensors}
In this appendix, we show some  nonezero components of Einstein's tensor and the energy momentum tensor of the aether field with the metric (\ref{metric}) and with the contraction of $R_{\mu\nu}=R^\sigma_{\;\mu\nu\sigma}$ are as following:
\begin{eqnarray}
&&G_{00}=\frac{(D-2)e^{2\phi-2\psi}}{2r^2}\Big[(D-3)(e^{2\psi}-1)+2r\psi'\Big],\\
&&G_{11}=\frac{D-2}{2r^2}\Big[(D-3)(1-e^{2\psi})+2r\phi'\Big],\\
&&G_{22}=e^{-2\psi}\Big[\frac{(D-3)(D-4)}{2}(1-e^{2\psi})+(D-3)r(\phi'-\psi')
+r^2(\phi''+\phi'^2-\phi'\psi')\Big],\\
&&G_{ii}=G_{22}\prod^{i-2}_{j=1}\sin^2\theta_j.
\end{eqnarray}
$T^{\ae}_{\mu\nu}$  are
\begin{eqnarray}
&&T^{\ae}_{00}=\frac{ b_1}{2}e^{-2\psi}h'^2-2\lambda h^2,\\
&&T^{\ae}_{01}=T^{\ae}_{10}=2e^\psi \lambda h\sqrt{e^{-2\phi}h^2\mp1},\\
&&T^{\ae}_{11}=-\frac{b_1}{2}e^{-2\phi}h'^2-2\lambda e^{2\psi}(e^{-2\phi}h^2\mp1),\\
&&T^{\ae}_{22}=\frac{b_1}{2}e^{-2(\phi+\psi)}r^2h'^2,\\
&&T^{\ae}_{ii}=T^{\ae}_{22}\prod^{i-2}_{j=1}\sin^2\theta_j.
\end{eqnarray}
The non-zero component of the divergence of the aether field strength $F_{\mu\nu}$ is
\begin{eqnarray}
\nabla ^\mu F_{\mu0}=-\frac{ 1}{r}e^{-2\psi}\Big[h'(-D+2+r\phi'+r\psi')-rh''\Big].
\end{eqnarray}

\end{document}